\documentclass{article}

\PassOptionsToPackage{numbers,sort&compress}{natbib}

\usepackage[final]{neurips}

\usepackage[utf8]{inputenc} %
\usepackage[T1]{fontenc}    %
\usepackage{hyperref}       %
\usepackage{url}            %
\usepackage{booktabs}       %
\usepackage{amsfonts}       %
\usepackage{nicefrac}       %
\usepackage{microtype}      %
\usepackage[table,xcdraw]{xcolor}         %
\usepackage{graphicx}
\usepackage{amsmath}
\DeclareRobustCommand{\swanicon}{\raisebox{-0.15em}{\includegraphics[height=1.1em]{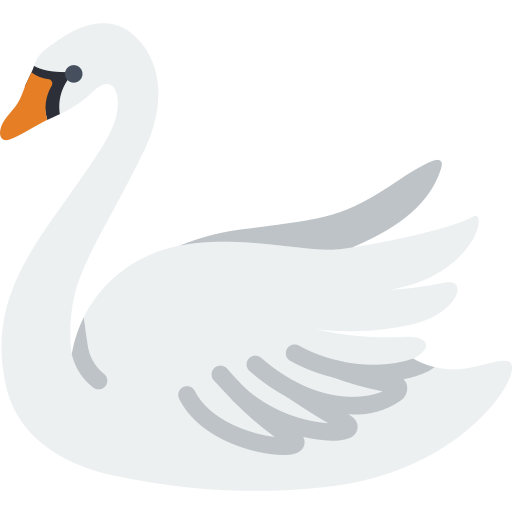}}}

\usepackage{newfloat}
\usepackage{listings}
\usepackage{pifont}

\usepackage{multirow}

\definecolor{darkorange}{rgb}{1.0, 0.55, 0.0}
\definecolor{mygreen}{RGB}{0,180,0}
\definecolor{myred}{RGB}{180,0,0}

\title{\swanicon\ SwanVoice: Expressive Long-Form Zero-Shot Speech Synthesis for Both Monologue and Dialogue}

\author{%
Ruiqi Li$^\textbf{1}$\thanks{Equal contribution}\quad
Yu Zhang$^\textbf{1}$\footnotemark[1]\quad
Changhao Pan$^\textbf{1,2}$\footnotemark[1]\quad
Ke Lei$^\textbf{1,2}$\quad
Xiang Yin$^\textbf{1}$\thanks{Corresponding Author}\quad
Cheng Yang$^\textbf{1}$\quad
\\
$^\textbf{1}$ByteDance, \ $^\textbf{2}$Zhejiang University\\
  \texttt{\{liruiqi.23,zhangyu.34,yinxiang.stephen\}@bytedance.com} \\
}

\begin{document}

\maketitle

\begin{abstract}
Zero-shot text-to-speech (TTS) has improved substantially for single-speaker synthesis, yet expressive long-form multi-speaker dialogue remains difficult. A common workaround is to synthesize each turn with a monologue TTS model and stitch the outputs together. This adds inference cost and often breaks acoustic consistency, conversational coherence, and affective continuity across turns. Recent dialogue TTS systems have begun to address this setting, but they still struggle to keep expressive coherence, controllable speaker switching, and monologue quality at the same time. We present SwanData-Speech and SwanVoice. SwanData-Speech builds monologue and dialogue corpora from in-the-wild audio, using Swan Forced Aligner for pause-aware word-level alignment and RobustMegaTTS3 for pronunciation-hard cases. Built on these data, SwanVoice is a zero-shot TTS model for 1--4 speakers, combining a 25 Hz VAE, raw-text conditioning with pause-aware symbols and pinyin substitution, and a flow-matching DiT with speaker-turn conditioning. Training starts from monologue speech, moves through mixed and real dialogue data, and then uses DiffusionNFT post-training with phone-level and speaker-similarity rewards. On SwanBench-Speech, SwanVoice obtains higher richness and hierarchy scores than all evaluated open-source baselines in both monologue and dialogue settings, while content accuracy remains the main limitation. Audio demos are available at \url{https://swanaigc.github.io/\#/swanvoice}.
\end{abstract}

\section{Introduction}
\label{sec: intro}

Recent advances in zero-shot text-to-speech (TTS) have made prompt-conditioned single-speaker synthesis increasingly reliable \cite{jiang2025megatts,chen2024f5,du2025cosyvoice,li2026indextts,ju2024naturalspeech,anastassiou2024seed,wang2023neural,guo2024fireredtts,wang2024maskgct,wang2025spark}. Many speech-generation applications, however, require more than single-speaker narration. Short-form dramas, podcasts, and similar settings need TTS systems that treat a multi-party conversation as one generation problem~\cite{zhang2025isdrama,ju2025mooncast}. The common workaround is to synthesize one turn at a time and concatenate the waveforms. This can preserve each speaker locally, yet adjacent turns may disagree in room response, background ambience, speaking intensity, or pause timing. The result sounds assembled rather than recorded as a scene. A dialogue model therefore has to model full conversations, not isolated turns.

Recent dialogue-capable TTS models have shown end-to-end two-speaker generation and controllable speaker switching \cite{ju2025mooncast,zhu2025zipvoice,xie2025fireredtts}. Long-form dialogue exposes failures that are less visible in short two-speaker generation: the acoustic environment should stay stable, speaker turns should remain separable even for similar voices, and affective continuity should carry across turns. At the same time, dialogue training should not degrade monologue synthesis. These failures are tightly coupled with data construction, since turn boundaries, pauses, and expressive labels shape turn control.

Architecturally, modern zero-shot TTS systems combine speech representations, neural vocoders, Transformer-based text/audio modeling, and a generative module such as diffusion or flow matching. They can be roughly divided into autoregressive (AR)~\cite{du2025cosyvoice,zhou2025indextts2} and non-autoregressive (NAR)~\cite{chen2024f5,jiang2025megatts} formulations. Several dialogue TTS models use AR designs~\cite{ju2025mooncast,xie2025fireredtts}. In long dialogue, however, language-model-style AR generation brings sequential latency and exposure-bias failures such as word skipping or repetition \cite{zhu2025zipvoice}. NAR generative modeling is a better fit here because it reduces sequential decoding latency and conditions on the full text and speaker-turn sequence at once.

Two bottlenecks are central to this paper.
1) \textbf{Dialogue data needs more than speaker labels.}
Expressive long-form synthesis needs speaker-consistent segments, pause-aware transcripts, quality filtering, and enough non-neutral speech to learn affective variation. These requirements interact: a speaker split error can corrupt turn control, while written-style punctuation can teach the model the wrong prosody.
2) \textbf{Dialogue training should not erase monologue ability.}
Many dialogue models start from a monologue model and fine-tune on dialogue data with speaker-switch labels~\cite{ju2025mooncast,zhu2025zipvoice}. This often improves turn control but can weaken monologue quality. The model also has to separate close voices, maintain a shared acoustic scene, and avoid pronunciation drift in long outputs.

We build \textbf{SwanData-Speech}, a pipeline for turning in-the-wild speech into monologue and dialogue training subsets. It is designed for sources such as podcasts, radio dramas, and film/TV content, where speakers, pauses, and acoustic conditions vary within long recordings. The pipeline includes: (i) a lightweight aligner, \textbf{Swan Forced Aligner}, for word-level timestamp alignment and pause-aware annotation; (ii) vocal separation and speaker segmentation modules built on existing methods; and (iii) quality and emotion filtering to retain clean expressive speech.

We then introduce \textbf{SwanVoice}, a zero-shot TTS model for 1--4 speakers. A 25 Hz VAE reduces the speech sequence length while preserving reconstruction quality. Raw text is kept as the main condition, with pause symbols and pinyin-substitution variants for pause control and Chinese pronunciation. The generator is a flow-matching DiT conditioned on speaker-turn IDs. SwanVoice is trained with a curriculum that moves from monologue speech to mixed and real conversational data, then post-trained with DiffusionNFT rewards for pronunciation robustness and speaker similarity.

\section{Data Processing Pipeline: SwanData-Speech}
\label{sec:data}

\subsection{Data Sources and Collection Scope}

SwanData-Speech begins with a raw collection drawn mainly from internal resources, together with selected open-source Chinese and English datasets for broader linguistic and acoustic coverage. The raw collection contains approximately 2.59 million hours of audio, including about 2.24 million hours of Chinese data and 0.35 million hours of English data. We process this collection into task-specific subsets: SwanVoice uses the filtered monologue and dialogue subsets produced by the pipeline, while the 80K-hour subset in \autoref{apx:forced-aligner} is reserved for training and evaluating Swan Forced Aligner.

SwanVoice uses raw text as the conditioning input. This preserves richer semantic information, but it also increases sparsity for rare and polyphonic characters. A training corpus cannot exhaust all characters, pronunciation variants, or corner cases such as Chinese–English code-switching. Replacing all text with pinyin would reduce part of the sparsity problem, but it would also reduce readability and make authoring less convenient.

We therefore construct RobustMegaTTS3, a pronunciation-hard synthetic subset later rendered with MegaTTS 3. We collect the full word list from GCIDE 0.54 and the Level-1 and Level-2 character lists from the Table of General Standard Chinese Characters \footnote{http://www.moe.gov.cn/publicfiles/business/htmlfiles/moe/cmsmedia/other/2013/7/other98742.zip}. An LLM (Qwen3-235B-A22B-Instruct-2507) then generates five example sentences per entry.

We also use the LLM to create 20K Chinese hard cases and 20K English hard cases, covering polyphonic-character disambiguation in context, erhua, tone sandhi, onomatopoeic characters, homographs with different pronunciations, noun–verb stress shift, and irregular spellings. Another 100K Chinese–English code-switching texts span 13 scenarios and roles to stress mixed-language synthesis.

To obtain accurate and standardized speech for these texts, we synthesize this portion of the audio with MegaTTS 3 \cite{jiang2025megatts}, a phoneme-pronunciation-based model. RobustMegaTTS3 supplies dictionary-level pronunciation knowledge for rare and ambiguous pronunciations.

\subsection{Pipeline Overview}

\begin{figure}[!htbp]
  \centering
    \includegraphics[width=0.9\textwidth]{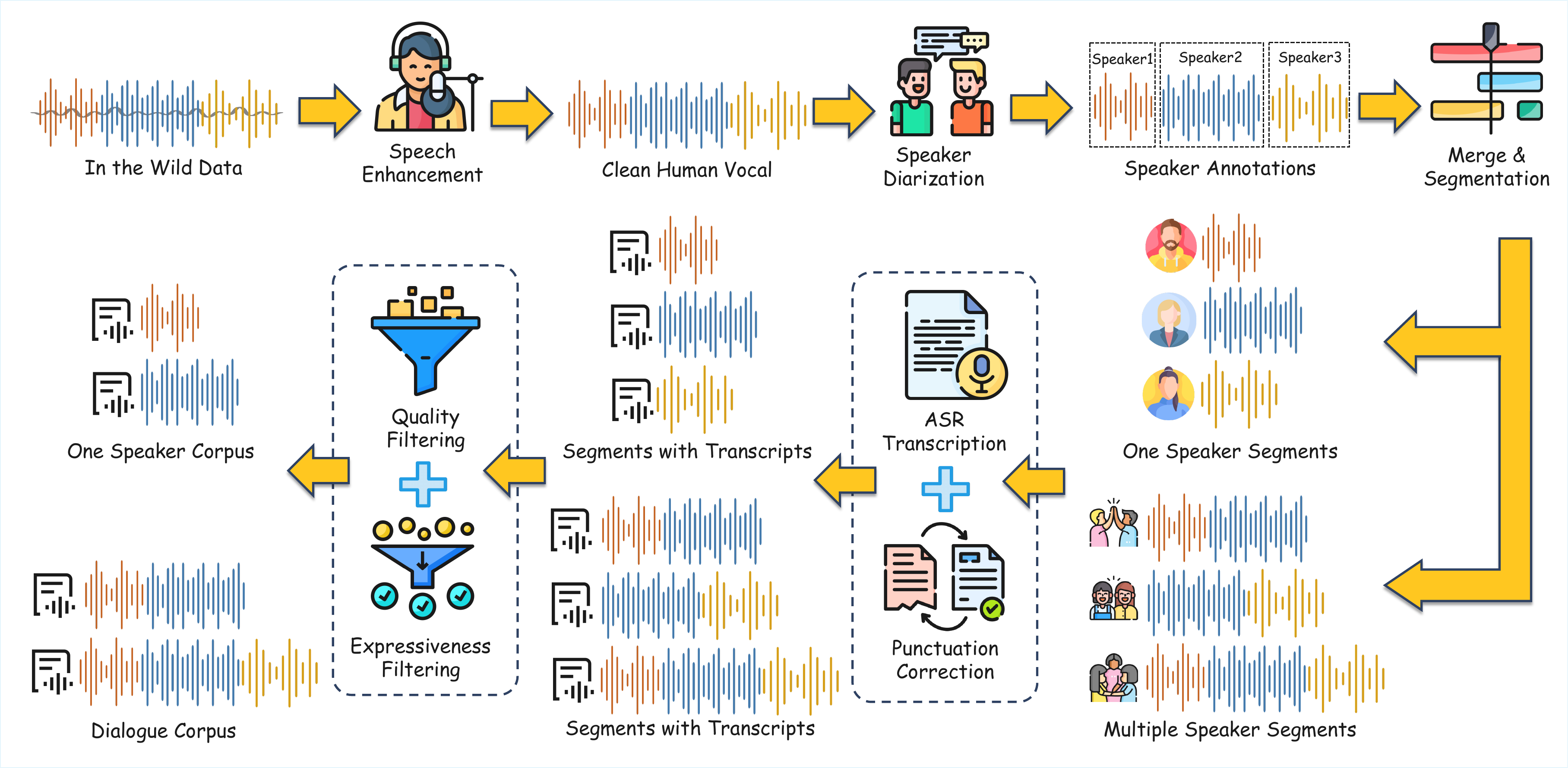}
  \caption{Hierarchical data processing pipeline}
  \label{fig:data_pipeline}
\end{figure}

The pipeline first applies speech enhancement and speaker diarization to raw audio. Based on speaker order, diarized segments are split into a monologue pool and a dialogue pool, and the two pools then go through ASR, punctuation refinement, and quality filtering separately. The output is two training datasets, one for monologue speech and one for dialogue conversations. We preserve the original sampling rate whenever possible during processing and resample all audio to 24 kHz only at the final stage. \autoref{fig:data_pipeline} summarizes the hierarchical processing pipeline.

\subsection{Segmentation and Speaker-Aware Processing}

\subsubsection{Speech Enhancement}

We apply a vocal separation tool \cite{ultimatevocalremovergui} to isolate the vocal component from all raw audio data. 

\subsubsection{Speaker Diarization}

Most raw recordings are long, often more than ten hours per sample, and may contain multiple speakers in arbitrary order. We therefore split them into shorter speaker-ordered segments.

We use the open-source 3D-Speaker toolkit \cite{chen20253d} for VAD, speaker embeddings, clustering, and diarization. It applies FSMN-Monophone VAD to split long audio into utterance-level chunks, then combines CAM++ \cite{wang2023campp} with spectral clustering for speaker-aware grouping.

After VAD and diarization, some segments are too short for stable training. We merge adjacent short segments from the same speaker when the silence between consecutive segments is at most 2 seconds. Segments shorter than 0.1 seconds, which are typically VAD artifacts, are removed, and each same-speaker merged sample is capped at 60 seconds.

For dialogue data, we merge consecutive multi-speaker segments up to 120 seconds. Each merged segment must contain 2--4 speakers, and no single silence interval may exceed 2 seconds. We use a sliding-window greedy merging strategy: starting from any monologue segment, a subsequent dialogue merge is kept as training data if it satisfies the constraints above. This partial overlap expands usable training data while preserving speaker order.

\subsection{Transcription and Alignment}
\subsubsection{ASR Transcription}

We use SenseVoice-Small \cite{an2024funaudiollm} for transcription and language identification, retaining only Chinese and English samples. Inverse text normalization (ITN) is disabled so that the model input stays closer to pronunciation; text normalization is left to a separate frontend model. Before pause correction, a small text Transformer restores punctuation for the transcribed text.

For dialogue utterances, we wrap the content of each speaker turn with special tokens of the form \texttt{<S\{id\}>} and \texttt{</S\{id\}>} to explicitly annotate the corresponding turn identity.

\subsubsection{Punctuation Correction}

The punctuation above is inferred from semantics. In conversational speech, however, semantic punctuation is often weakly correlated with actual pauses. A model trained on such text may learn to ignore punctuation and rely on dataset statistics for pause behavior, which leads to poor prosody in synthesized dialogue, especially around turn boundaries.

We revise punctuation in the transcribed text to better match acoustic pause patterns. A pretrained forced aligner first aligns the audio with the transcription and assigns a timestamp to each character. Pauses are then defined by the time gap between consecutive characters. Pauses shorter than 0.08 s are ignored. For pauses between 0.08 s and 0.18 s, we insert \texttt{<|sp|>}. For pauses between 0.18 s and 0.45 s, we use a comma. For pauses longer than 0.45 s, we use a period, exclamation mark, or question mark, depending on the original punctuation before correction; the default is a period. If punctuation appears where no pause is observed, it is removed. If a pause is observed without punctuation, punctuation is inserted. The aligner design and evaluation are reported in \autoref{apx:forced-aligner}.

\subsection{Data Filtering}

We score all audio samples with the non-intrusive DNSMOS metric \cite{reddy2021dnsmos}. PESQ \cite{rix2001pesq} and STOI \cite{zezario2020stoi} are originally intrusive metrics, but we use the non-intrusive PESQ and STOI models from torchaudio-SQUIM \cite{kumar2023torchaudiosquim} to score the full corpus.

After the initial filtering stage, emotion2vec+ \cite{ma2024emotion2vec} classifies the emotion of each sample and produces a confidence score. High-confidence non-neutral samples define the high-expressiveness subset.

\section{Method: SwanVoice}
\label{sec: method}

\begin{figure}[!htbp]
  \centering
    \includegraphics[width=1.0\textwidth]{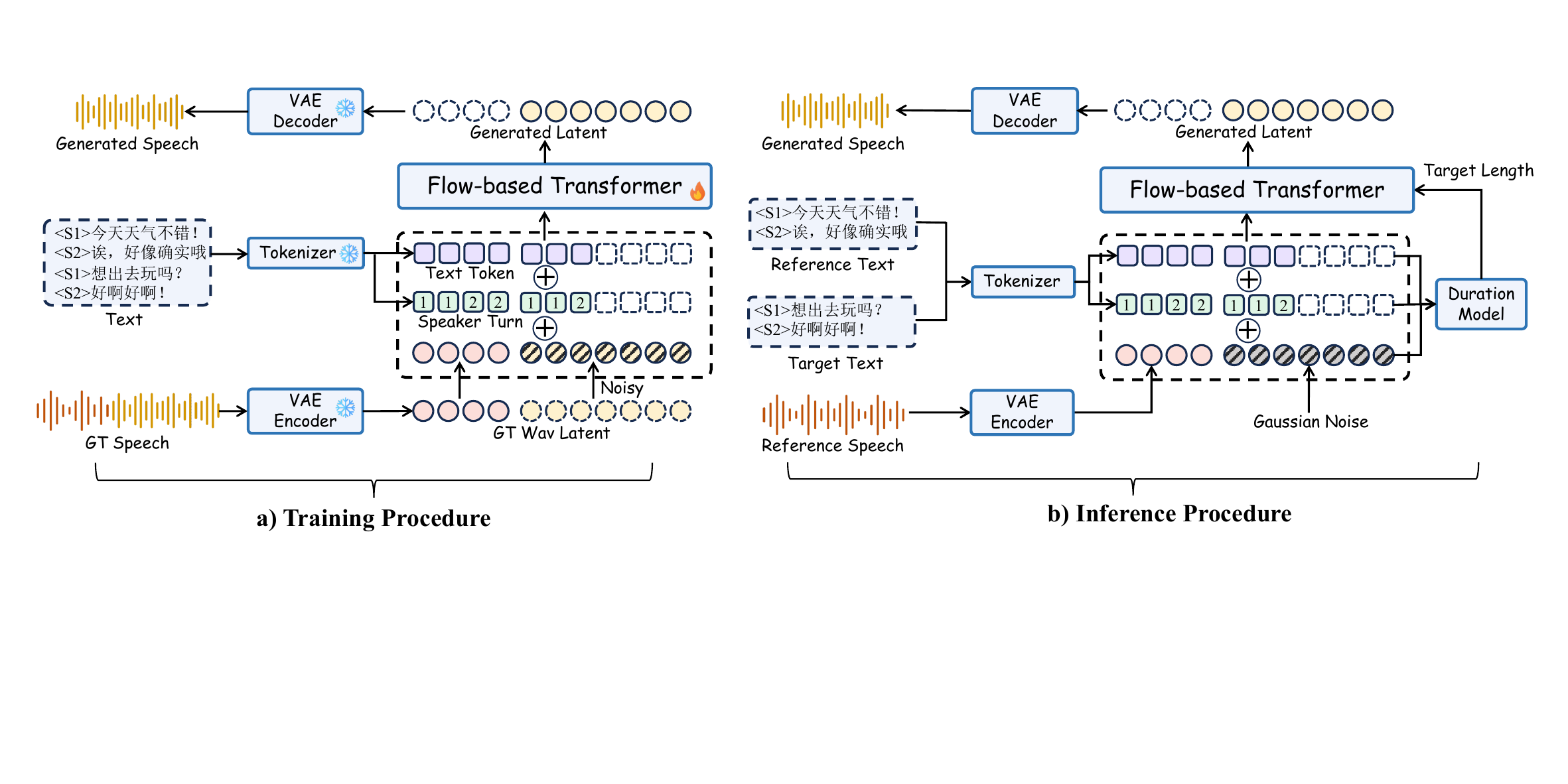}
  \caption{Overall training and inference procedure of SwanVoice.}
  \label{fig:arch}
\end{figure}

\subsection{VAE}

Given a speech waveform $s$, a variational encoder $E$ maps $s$ to a latent representation $z$, and a waveform decoder $D$ reconstructs the signal as $\hat{s}=D(z)=D(E(s))$.
To reduce computational cost and ease subsequent speech--text alignment, $E$ temporally downsamples the input waveform by a factor of $d$.
Architecturally, $E$ follows the design in \citet{ji2024wavtokenizer}, while $D$ is built upon HiFi-GAN \citep{kong2020hifi}.
To capture high-frequency details and improve perceptual fidelity, we train with a set of adversarial discriminators, including the multi-period discriminator (MPD), multi-scale discriminator (MSD), and multi-resolution discriminator (MRD) \citep{kong2020hifi,jang2021univnet}.
The overall training objective is
\begin{equation}
\mathcal{L}=\mathcal{L}_{\mathrm{rec}}+\mathcal{L}_{\mathrm{KL}}+\mathcal{L}_{\mathrm{Adv}},
\end{equation}
where $\mathcal{L}_{\mathrm{rec}}=\|\Phi(s)-\Phi(\hat{s})\|_2^2$ denotes the spectrogram-domain reconstruction loss computed by a feature extractor $\Phi$, $\mathcal{L}_{\mathrm{KL}}$ is a lightly-weighted KL regularizer as in \citet{rombach2022high}, and $\mathcal{L}_{\mathrm{Adv}}$ is an LSGAN-style adversarial loss \citep{mao2017least}.
The compression rate is 25 latent frames per second.

\subsection{Tokenizer}

We use the CosyVoice tokenizer \cite{du2025cosyvoice} and feed raw text directly to the model. The text is tokenized by a BPE-based tokenizer, removing the need for a separate grapheme-to-phoneme (G2P) frontend. This simplifies preprocessing while allowing the model to learn context-dependent pronunciations end to end. For Chinese, the tokenizer provides a one-to-one character-level encoding, which prevents a single token from carrying an excessively long pronunciation and reduces sparse corner cases.

We add a dedicated pause token, \texttt{<|sp|>}, so the model can learn natural pausing patterns from text. For Chinese pronunciation control, the tokenizer vocabulary is augmented with 1,549 pinyin syllable combinations. During training, we randomly replace a subset of Chinese characters with pinyin forms extracted by \texttt{pypinyin}. This improves robustness to pronunciation variation. At inference time, pinyin hints can enforce the desired pronunciation of a character, which is useful for polyphonic characters and certain Northern Chinese dialect pronunciations.

For speaker annotation, we add a speaker-turn label sequence with the same length as the text-token sequence. Each label indicates the speaker identity of the corresponding token. During text preprocessing, each speaker's content is wrapped with turn-specific tags \texttt{<S\{id\}>} and \texttt{</S\{id\}>}. The speaker label sequence is constructed by detecting these tags and assigning the corresponding speaker ID to each token span.

\subsection{Flow-based Transformer}

As shown in Figure~\ref{fig:arch}(a), the diffusion transformer (DiT) pads the text-token sequence and speaker-turn embeddings to the temporal resolution of the waveform latent sequence. Instead of concatenating these heterogeneous conditions with the speech latent at the input, we first pass the padded text and turn representations through a lightweight Transformer stack. The model can therefore form text-side and turn-side features before they interact with the speech representation. Compared with naive early concatenation, this strategy improves in-context conditioning on the speech input~\cite{chen2024f5}.

The ground-truth waveform latent is constructed from a complete utterance. For monologue data, multiple short utterances from the same speaker may be concatenated into a longer sentence-level training example to improve long-form modeling. The waveform is encoded by the VAE into a latent sequence $\mathbf{z}^{\star}$. We randomly split $\mathbf{z}^{\star}$ into two contiguous parts: the first part is used as the \emph{reference} segment, and the second part is the \emph{target} segment to be generated. For dialogue data, the reference segment is required to contain at least a short span of speech from every speaker. Gaussian noise is injected into the target latent. The noised target, clean reference latent, processed text, and speaker-turn conditions are then fed into a flow-based Transformer. The Transformer is implemented as a deep stack of self-attention blocks and estimates the vector field over the latent trajectory.

We use RMSNorm~\citep{zhang2019root} throughout the network and add AdaLN-based global adapters~\citep{peebles2023scalable} to stabilize optimization and preserve long-form consistency in speaker timbre and recording conditions.

Following the standard flow-matching formulation, the model is trained to predict the velocity field between a noise sample and the clean target latent:
\begin{equation}
\label{eq:flow_loss_revised}
\mathcal{L}_{\mathrm{flow}}
=
\mathbb{E}_{t \sim \mathcal{U}(0,1),\, \mathbf{z}^{\star} \sim p_{\mathrm{data}},\, \boldsymbol{\epsilon} \sim \mathcal{N}(\mathbf{0}, \mathbf{I})}
\left[
\left\|
\mathbf{u}_{\theta}(\mathbf{z}_{t}, t, \mathbf{c})
-
\left(\mathbf{z}^{\star} - \boldsymbol{\epsilon}\right)
\right\|_{2}^{2}
\right],
\end{equation}
where
\begin{equation}
\mathbf{z}_{t} = (1-t)\boldsymbol{\epsilon} + t\mathbf{z}^{\star},
\end{equation}
and $\mathbf{c}$ denotes the full conditioning information, including the processed text tokens, speaker-turn embeddings, and the reference speech latent used for conditioning.

\subsection{Curriculum Learning}

Training directly on conversational data from scratch often produces unintelligible speech. The main difficulty is learning speech-text alignment from spoken conversations with multiple speakers, while still preserving strong monologue performance. We therefore use a three-stage curriculum that gradually moves from monologue data to real conversational data.

\textbf{1) Monologue pretraining.}
We first train the model from scratch on monologue speech data. This stage uses approximately 2 million hours of monologue speech covering both Chinese and English. It establishes the basic synthesis ability, including high-fidelity acoustic modeling and reliable speech--text alignment. Starting the later stages from this pretrained model avoids many of the audio-quality and pronunciation failures that appear when training directly on complex conversational data.

This stage is also augmented with the pronunciation-hard and code-switching synthetic cases described in Section~\ref{sec:data}. These cases are difficult to collect at scale, and phoneme-based synthesis covers pronunciations that are rare in crawled speech.

\textbf{2) Mixed conversational training.}
In the second stage, the pretrained monologue model is trained on monologue data together with concatenated 2--4-speaker conversational data. Since speaker diarization is imperfect, directly using real conversational data can make speaker transitions difficult to learn. The concatenated data provides an intermediate step in which the model learns to assign the correct speaker identity to each turn. Conversational examples are sampled more often than their raw-hour share so the model sees speaker switches frequently, while monologue examples remain in the mixture to prevent monologue degradation.

\textbf{3) SFT training.}
In the third stage, the model is trained on monologue data together with real 2--4-speaker conversational data. By this point, it already has stable speaker-switching ability, so real conversational data can be used to learn higher-level dialogue consistency, including recording-environment consistency and emotional coherence. Monologue examples remain in the mixture to protect monologue performance. The real conversational data mainly comes from movies, TV dramas, and podcasts, which expose the model to richer affective and conversational variation.

\subsection{Post Training}

After supervised training, the DiT-based TTS model still makes predictable errors: difficult words may be misread, and prompt speaker identity can drift. We address these errors with a post-training stage that optimizes model-generated samples against pronunciation and timbre rewards. Since usable reward models are available in our setting, we use online reinforcement learning. Instead of introducing an additional value model, we use a value-free optimization strategy and instantiate it with DiffusionNFT~\cite{zheng2025diffusionnft}, which matches the flow-matching backbone. The rewards target phone-level consistency and speaker similarity, not recording-environment consistency or expressiveness.

Flow-GRPO \cite{liu2025flow} is an early attempt to bring online RL to flow-matching models. It converts the deterministic ODE sampling process into an equivalent SDE for stochastic exploration and uses a denoising-reduction strategy to lower training cost. DiffusionNFT is simpler for this setting: it performs policy optimization on the forward process through the flow-matching objective, addresses the forward-inconsistency issue of reverse-process RL, allows arbitrary black-box solvers, and only requires final clean samples with rewards rather than the full latent trajectory. DiffusionNFT also reports better efficiency than Flow-GRPO in head-to-head comparisons.

\subsubsection{Reward Models}

The reward has two components: an ASR-based robustness reward for intelligibility and recognition errors, and a speaker-similarity reward for timbre preservation. Differentiable ASR-based optimization is possible in principle, but it would complicate the training pipeline and is not needed here. We use a reward-driven online RL formulation in which the model is updated from sampled utterances and their rewards without differentiating through the recognizers.

The first reward is the \textbf{phone consistency reward} $r_{\mathrm{phone}}$, which measures how well the generated speech matches the target text at the phoneme and tone levels. We apply an external phone recognizer to $\hat{x}$ and compare the resulting phonetic sequence with the phonetic realization implied by $y$, yielding a normalized score in $[0, 1]$.

We remove punctuation and silence symbols on both sides, and merge each phone-tone pair into a single token, e.g., $u_j = \texttt{phone}_j\_\texttt{tone}_j$. Let $\mathbf{u}^{\mathrm{ref}}$ and $\mathbf{u}^{\mathrm{hyp}}$ denote the resulting reference and predicted token sequences. The resulting WER and phone reward are
\begin{align}
\mathrm{WER}(\mathbf{u}^{\mathrm{ref}}, \mathbf{u}^{\mathrm{hyp}})
&=
\frac{S + D + I}{\max(1, |\mathbf{u}^{\mathrm{ref}}|)},\\
r_{\mathrm{phone}}
&=
\exp\!\big(-\mathrm{WER}(\mathbf{u}^{\mathrm{ref}}, \mathbf{u}^{\mathrm{hyp}})\big),
\end{align}
where $S$, $D$, and $I$ are the numbers of substitutions, deletions, and insertions, respectively.
We use a phone-based recognizer rather than the character- or word-based recognizers often used in ASR-derived rewards because the objective here is pronunciation accuracy and polyphonic-character disambiguation, especially in Chinese, rather than exact character identity.

The second reward is a \textbf{speaker similarity reward} $r_{\mathrm{sim}}$\footnote{https://github.com/microsoft/UniSpeech/tree/main/downstreams/speaker\_verification}, which compares the generated speech with the reference prompt in a pretrained speaker-embedding space:
\begin{equation}
r_{\mathrm{sim}}(\hat{x}, x^{\mathrm{ref}}) =
\cos\!\big(f_{\mathrm{spk}}(\hat{x}), f_{\mathrm{spk}}(x^{\mathrm{ref}})\big),
\end{equation}
where $f_{\mathrm{spk}}(\cdot)$ is a frozen speaker encoder. We aggregate the two rewards as
\begin{equation}
r = \frac{1}{2}\big(r_{\mathrm{phone}} + r_{\mathrm{sim}}\big),
\end{equation}
which is the default setting in our experiments. The framework also supports a weighted sum of multiple rewards when deployment priorities require different trade-offs.

\subsubsection{DiffusionNFT-style Policy Optimization}

For each prompt, we draw multiple candidates from $\pi_{\mathrm{old}}$ and compute their rewards. A prompt-wise advantage is formed by subtracting a within-prompt baseline:
\begin{equation}
A_i = r_i - \bar{r}, \ \bar{r} = \frac{1}{K}\sum_{j=1}^{K} r_j,
\end{equation}
where $K$ is the number of sampled candidates for the same condition. We clip the advantage and map it into a soft preference weight:
\begin{equation}
\tilde{A}_i = \mathrm{clip}(A_i, -A_{\max}, A_{\max}), \quad
w_i = \mathrm{clip}\!\left(\frac{\tilde{A}_i}{2A_{\max}} + \frac{1}{2}, 0, 1\right).
\end{equation}

Let $v_{\theta}(z_t, t, c)$, $v_{\mathrm{old}}(z_t, t, c)$, and $v_{\mathrm{ref}}(z_t, t, c)$ denote the denoising predictions of the online, old, and reference policies, respectively, under latent state $z_t$, timestep $t$, and condition $c=(y, x^{\mathrm{ref}})$. Following the NFT-style update rule, we construct positive and implicit-negative denoising branches:
\begin{align}
v_i^{+} &= \beta_{\mathrm{NFT}} v_{\theta} + (1-\beta_{\mathrm{NFT}})\,\mathrm{sg}\!\left(v_{\mathrm{old}}\right),\\
v_i^{-} &= (1+\beta_{\mathrm{NFT}})\,\mathrm{sg}\!\left(v_{\mathrm{old}}\right) - \beta_{\mathrm{NFT}} v_{\theta},
\end{align}
where $\mathrm{sg}(\cdot)$ denotes stop-gradient and $\beta_{\mathrm{NFT}}$ controls the interpolation strength. These predictions are converted to denoised latent estimates for the non-prompt region. The online policy is optimized to prefer the positive branch when $w_i$ is large and the implicit negative branch when $w_i$ is small. The objective can be written as
\begin{equation}
\mathcal{L}_{\mathrm{NFT}}
=
\mathbb{E}_{i}\left[
\frac{w_i}{\beta_{\mathrm{NFT}}}\,\ell\!\left(\hat{z}_{0,i}^{+}, z_{0,i}\right)
+
\frac{1-w_i}{\beta_{\mathrm{NFT}}}\,\ell\!\left(\hat{z}_{0,i}^{-}, z_{0,i}\right)
\right],
\end{equation}
where $\ell(\cdot,\cdot)$ denotes the masked denoising loss on the generated target segment. To prevent the policy from drifting too far from the pretrained model, we add a reference-policy regularizer:
\begin{equation}
\mathcal{L}
=
\mathcal{L}_{\mathrm{NFT}}
+
\lambda_{\mathrm{ref}} \mathcal{L}_{\mathrm{ref}}, \quad
\mathcal{L}_{\mathrm{ref}}
=
\mathbb{E}\big[\|v_{\theta} - \mathrm{sg}(v_{\mathrm{ref}})\|_2^2\big].
\end{equation}
This reference regularization preserves the speech quality and robustness inherited from supervised pretraining while still allowing reward-driven adaptation.

For post-training data, we collected 3K audio samples of real human conversations, transcribed them into text, and corrected the pause annotations. The post-training objective explicitly optimizes only phone-level WER and speaker similarity. In qualitative inspection, the resulting model also shows better recording-environment consistency and stronger expressiveness, which we treat as side effects.

\subsection{Inference Procedure}

As shown in Figure~\ref{fig:arch}(b), inference takes a reference speech segment and a target text sequence as input. The model synthesizes the target linguistic content while preserving the speaker identity and speaking style of the reference speech. We transcribe the reference speech with SenseVoice-Small~\cite{an2024funaudiollm} to obtain speaker-specific reference text. The target duration is estimated with a simple speaking-rate heuristic for each speaker in the reference speech. We also use sway sampling~\cite{chen2024f5}, which encourages the model to capture coarse speech contours in the early generation stage and refine fine-grained details later. The speech-text alignment is therefore largely determined by the first few denoising steps. Finally, the VAE decoder converts the target latent into a waveform.

We also introduce a staircase classifier-free guidance (CFG) strategy. It uses two guidance scales and three conditioning variants: a null condition, a full condition, and a text-only condition. The guided prediction is defined as
\[
\tilde{v}_t
=
v_{\emptyset}
+
\omega_{\mathrm{text}}\bigl(v_{\mathrm{text}} - v_{\emptyset}\bigr)
+
\omega_{\mathrm{ref}}\bigl(v_{\mathrm{full}} - v_{\mathrm{text}}\bigr),
\]
where \(\omega_{\mathrm{text}}\) and \(\omega_{\mathrm{ref}}\) denote the guidance scales for textual content and reference-dependent speaker/style information. The staircase formulation separates content guidance from reference guidance, allowing the two effects to be controlled independently during inference. Increasing \(\omega_{\mathrm{ref}}\) moves the output toward the reference timbre and style without changing text guidance.

\section{Experiments}
\label{sec: exp}

\subsection{Implementation Details}

The main SwanVoice model has 2 billion parameters. Monologue pretraining uses 64 A100 GPUs for 500k steps, mixed conversational training uses 32 A100 GPUs for 600k steps, and supervised fine-tuning (SFT) uses 32 A100 GPUs for 300k steps. Post-training uses 8 A100 GPUs for 50 epochs.

\subsection{Evaluation Metrics}
Following the evaluation protocol of SwanBench-Speech~\cite{pan2026swanbench}, we evaluate each model along three axes: acoustics, semantics, and expressiveness.

\paragraph{Acoustics}
For acoustics, we report timbre consistency, reverb consistency, and sound fidelity.
Timbre consistency is measured with segment-based speaker similarity, computed as the average similarity of speaker embeddings~\footnote{\url{https://huggingface.co/docs/transformers/en/model_doc/unispeech-sat}} across segments.
Reverb consistency follows the same idea: we compute the standard deviation of SRMR~\footnote{\url{https://github.com/jfsantos/SRMRpy}} values within sliding windows to measure the stability of the synthesized acoustic environment.
Sound fidelity is measured by SQUIM-PESQ through the official \texttt{torchaudio} interface as a non-intrusive, reference-free metric~\footnote{\url{https://docs.pytorch.org/audio/main/tutorials/squim_tutorial.html}}.

\paragraph{Semantics} 
For semantics, we evaluate content error rate and prosodic coherence.
Content errors are measured by Character Error Rate (CER) on Chinese datasets and Word Error Rate (WER) on English datasets. Both are computed with FunASR-Nano~\cite{an2024funaudiollm} as the ASR model and \texttt{JiWER} as the calculation backend.
For prosody, we use SpeechJudge~\cite{zhang2025speechjudge}, a Qwen2.5-Omni model fine-tuned for audio quality assessment. Prosodic coherence is rated on a 1.0--5.0 scale, where 1 means poor coherence and 5 means excellent coherence.

\paragraph{Expressiveness} We evaluate expressiveness from two views: sentence-level expressive richness and paragraph-level expressive hierarchy for long-form speech.
Because MOS prediction networks can correlate poorly with human perception~\cite{minixhofer2025ttsds2}, we use an MLLM-as-a-judge protocol with a large audio language model as the evaluator.
For expressive richness, the audio waveform is segmented into non-overlapping 10-second chunks $\{c_i\}_{i=1}^M$. The evaluator assigns an expressiveness score $s_i$ to each chunk $c_i$, and the final richness score is the arithmetic mean:
$\text{Score}_{\text{rich}} = (\sum_{i=1}^{M} s_i) / M$. 
For expressive hierarchy, the full audio sequence is fed into the evaluator, which scores the speech along three dimensions: Emotional Variation, Vocal Dynamics, and Scene Appropriateness.
We use Gemini-3-Pro as the evaluator and report both scores on a 1--5 scale, where 1 is poor and 5 is excellent. The evaluator is given only the audio and a fixed scoring rubric, without system names, and samples from different systems are evaluated in a randomized order.

\begin{table*}[t]
    \caption{
        \textbf{Evaluation results of long-form TTS models across multi-dimensional metrics.} Metrics cover Acoustics (Timbre/Reverb Consistency, Sound Fidelity), Semantics (Content Error, Prosodic Coherence), and Expressiveness (Richness, Hierarchy).
        The best and second-best results are marked in \textbf{bold} and \underline{underlined}, respectively, for each metric.
    }
    \label{tab:mono_res}
    \vspace{2pt}
    \centering
  \resizebox{\linewidth}{!}{
        \begin{tabular}{@{}lccccccc@{}}
        \toprule
        \multicolumn{1}{c|}{}                                 & \multicolumn{3}{c|}{\textbf{Acoustics}}                                    & \multicolumn{2}{c|}{\textbf{Semantics}}                     & \multicolumn{2}{c}{\textbf{Expressiveness}} \\ \cmidrule(l){2-8} 
        \multicolumn{1}{c|}{\multirow{-2}{*}{\textbf{Model}}} & \textbf{Timbre($\uparrow$)} & \textbf{Reverb($\downarrow$)} & \multicolumn{1}{c|}{\textbf{Sound Fidelity($\uparrow$)}}              & \textbf{Content Error($\downarrow$)} & \multicolumn{1}{c|}{\textbf{Prosody($\uparrow$)}}              & \textbf{Richness($\uparrow$)}          & \textbf{Hierarchy($\uparrow$)}        \\ \midrule
        \rowcolor[HTML]{EFEFEF} 
        \multicolumn{8}{c}{\cellcolor[HTML]{EFEFEF}\textit{\textbf{Open-Source Models}}}                    \\ \midrule
        \multicolumn{1}{c|}{CosyVoice-2}   & {0.93} & {2.37} & \multicolumn{1}{c|}{3.58} & {0.106} & \multicolumn{1}{c|}{2.81}   & {2.02}  & {2.59}  \\
        \multicolumn{1}{c|}{CosyVoice-3}   & 0.93  & 2.73     & \multicolumn{1}{c|}{3.80}     & {0.077}    & \multicolumn{1}{c|}{3.26}  &  {2.64}  & {2.47}  \\
        \multicolumn{1}{c|}{FishSpeech}     & {0.93} &   {2.00}    & \multicolumn{1}{c|}{\textbf{4.09}}    &  {\textbf{0.066}}       & \multicolumn{1}{c|}{\textbf{3.77}}   &  {2.37} & {2.90}                      \\
        \multicolumn{1}{c|}{F5TTS}       &  {0.92}    &  {2.12}   & \multicolumn{1}{c|}{2.60}   &    {0.085}    & \multicolumn{1}{c|}{2.87}        & {2.77}   &  {2.97}    \\
        \multicolumn{1}{c|}{GLM-TTS}   &  \textbf{0.94}   &  \textbf{1.64}    & \multicolumn{1}{c|}{\underline{3.90}}  &   {0.074}  & \multicolumn{1}{c|}{3.28} & {1.57} & {2.39}                     \\
        \multicolumn{1}{c|}{IndexTTS-2}    &  0.93   & \underline{1.77} & \multicolumn{1}{c|}{2.78}   &  {0.077}  & \multicolumn{1}{c|}{3.63}    &  3.32     &  2.94          \\
        \multicolumn{1}{c|}{MegaTTS-3}   &  {0.93}  &   {2.07}   & \multicolumn{1}{c|}{3.52} &   {\underline{0.072}}    & \multicolumn{1}{c|}{3.22}   &  {2.40}    & {3.01}   \\
        \multicolumn{1}{c|}{SparkTTS}   & {0.92} & {2.04} & \multicolumn{1}{c|}{3.53}   &  {0.314}  & \multicolumn{1}{c|}{2.35}  & {2.23}   & {2.22}        \\
        \multicolumn{1}{c|}{VibeVoice}     &   {0.92}  &  {2.45}  & \multicolumn{1}{c|}{3.47}     &    {0.092}   & \multicolumn{1}{c|}{\underline{3.75}} & \underline{3.42}  &  \underline{3.06}  \\
        \multicolumn{1}{c|}{ZipVoice}      &  0.89   & 2.10     & \multicolumn{1}{c|} {3.53}  & 0.213   & \multicolumn{1}{c|}{2.97}   & {2.11}   & {2.05}       \\
        \rowcolor[HTML]{FFFC9E} 
        \multicolumn{1}{c|}{Average}  &  0.92  & 2.13   & \multicolumn{1}{c|}{3.48} &    {0.12}   & \multicolumn{1}{c|}{3.19} &   2.49    &   2.66
        \\ \midrule
        \multicolumn{1}{c|}{\textbf{SwanVoice}}                                           &   \underline{0.93}    &  2.06   & \multicolumn{1}{c|}{3.60}  &    0.172   & \multicolumn{1}{c|}{3.56}       &   \textbf{3.81}    &  \textbf{3.62}      \\
        \bottomrule
        \end{tabular}
    }
\end{table*}

\subsection{Baselines}

For monologue generation, we compare with ten open-source models: ZipVoice~\cite{zhu2025zipvoice}, SparkTTS~\cite{wang2025spark}, CosyVoice2-0.5B~\cite{du2024cosyvoice2}, CosyVoice3-0.5B~\cite{du2025cosyvoice}, GLM-TTS~\cite{cui2025glm}, MegaTTS3~\cite{jiang2025megatts}, IndexTTS2~\cite{zhou2025indextts2}, FishSpeech-1.5~\cite{liao2024fish}, F5TTS~\cite{chen2024f5}, and VibeVoice~\cite{peng2025vibevoice}.

For dialogue generation, we compare with six open-source long-form models: ZipVoice-Dialog~\cite{zhu2025zipdialog}, MoonCast~\cite{ju2025mooncast}, MOSS-TTSD~\cite{zhao2025moss}, FireRedTTS2~\cite{xie2025fireredtts}, VibeVoice, and \mbox{SoulX-Podcast}~\cite{xie2025soulx}.

\subsection{Zero-Shot Monologue TTS}

Table~\ref{tab:mono_res} reports results on the Expressive Challenge subset of SwanBench-Speech. SwanVoice reaches 3.81 in richness and 3.62 in hierarchy, higher than all evaluated open-source baselines. Relative to VibeVoice, the strongest baseline on these two metrics, the gains are 0.39 and 0.56 points. The model is not the best on content error, but it keeps 0.93 timbre consistency, 3.60 sound fidelity, and 3.56 prosodic coherence, all at or above the open-source average.

\subsection{Zero-Shot Dialogue TTS}

\begin{table*}[t]
    \caption{
        \textbf{Results of dialogue generation models across SwanBench-Speech metrics}.
        The best and second-best results are marked in \textbf{bold} and \underline{underlined}, respectively, for each metric.
    }
    \label{tab:two_res}
    \centering
    \footnotesize
  \resizebox{\linewidth}{!}{
        \begin{tabular}{@{}lccccccc@{}}
        \toprule
        \multicolumn{1}{c|}{}                                 & \multicolumn{3}{c|}{\textbf{Acoustics}}                                    & \multicolumn{2}{c|}{\textbf{Semantics}}                     & \multicolumn{2}{c}{\textbf{Expressiveness}} \\ \cmidrule(l){2-8} 
        \multicolumn{1}{c|}{\multirow{-2}{*}{\textbf{Model}}} & \textbf{Timbre($\uparrow$)} & \textbf{Reverb($\downarrow$)} & \multicolumn{1}{c|}{\textbf{Sound Fidelity($\uparrow$)}}              & \textbf{Content Error($\downarrow$)} & \multicolumn{1}{c|}{\textbf{Prosody($\uparrow$)}}              & \textbf{Richness($\uparrow$)}          & \textbf{Hierarchy($\uparrow$)}          \\ \midrule
        \rowcolor[HTML]{EFEFEF} 
        \multicolumn{8}{c}{\cellcolor[HTML]{EFEFEF}\textit{\textbf{Open-Source Models}}}                                                                                                                                                              \\ \midrule
        \multicolumn{1}{c|}{FireRedTTS-2}   & {\underline{0.91}} & 3.54 & \multicolumn{1}{c|}{2.54}      & {0.148} & \multicolumn{1}{c|}{2.93}   & {2.52} & {2.65} \\
        \multicolumn{1}{c|}{MoonCast}       & {0.90} & 3.29  & \multicolumn{1}{c|}{2.60}    &  {0.284}  & \multicolumn{1}{c|}{2.93}    & {2.42} & {2.54}       \\
        \multicolumn{1}{c|}{MOSS-TTSD}      &  {0.89} & {3.52}  & \multicolumn{1}{c|}{2.83}    & {0.227}   & \multicolumn{1}{c|}{2.57}    & {3.04}     & {2.86}     \\
        \multicolumn{1}{c|}{SoulX-Podcast}  &  {\textbf{0.92}} & {3.23}  & \multicolumn{1}{c|}{\textbf{3.98}}    & {\textbf{0.101}}  & \multicolumn{1}{c|}{\textbf{3.89}}  & {2.80}  & \underline{3.15}        \\
        \multicolumn{1}{c|}{VibeVoice}      & {0.89} & {\textbf{2.09}}  & \multicolumn{1}{c|}{2.75}    &  {0.204}   & \multicolumn{1}{c|}{3.00}    & \underline{3.09}  & {2.83}                     \\
        \multicolumn{1}{c|}{ZipVoice-Dialog}     & {0.90} & {3.49} & \multicolumn{1}{c|}{2.48}   & \underline{0.116}   & \multicolumn{1}{c|}{3.46}   & {2.88}  & {2.93}      \\
        \rowcolor[HTML]{FFFC9E} 
        \multicolumn{1}{c|}{Average}  &  0.90  &   3.19   & \multicolumn{1}{c|}{2.86} &    {0.180}  & \multicolumn{1}{c|}{3.13} &  2.79   & 2.83              \\ \midrule
        \multicolumn{1}{c|}{\textbf{SwanVoice}}        &    \textbf{0.92}  &  \underline{3.02}   & \multicolumn{1}{c|}{\underline{3.77}}     &  {0.145} & \multicolumn{1}{c|}{\underline{3.70}}   &  \textbf{3.62}  &  \textbf{3.71} \\
        \bottomrule
        \end{tabular}
    }
\end{table*}

For dialogue, SwanVoice reaches 3.62/3.71 on richness/hierarchy, 0.53/0.56 points higher than the strongest baselines. Content error is below the baseline average but not the best in the table, and the demo page further includes 3--4-speaker cases.

\section{Conclusion}
\label{sec: con}

SwanVoice treats long-form dialogue as a full-context generation problem rather than a sequence of isolated turns. In our experiments, this matters most for expressiveness: SwanVoice obtains higher richness and hierarchy scores than all evaluated open-source baselines in both monologue and dialogue settings. The data pipeline contributes directly to this behavior. Speaker-aware segmentation, pause-aware alignment, pronunciation hard cases, and emotion-based filtering each address a failure mode that becomes audible in long speech.

The current model still has clear limitations. Content accuracy remains weaker than the best baselines in several settings, and speaker switching can still fail when the speakers are acoustically close or when the prompt is short. These errors suggest three directions for improvement: pronunciation control, alignment and pause modeling, and more robust speaker-turn conditioning. Future work should therefore focus on making long-form speech generation more reliable.

\newpage
\bibliographystyle{neurips}
\bibliography{custom}

\newpage
\appendix
\begin{center}{\bf {\LARGE Appendices} }
\end{center}
\begin{center}{\bf {\Large \swanicon\ SwanVoice: Expressive Long-Form Zero-Shot Speech Synthesis for Both Monologue and Dialogue} \linebreak}
\end{center}

\section{Swan Forced Aligner} \label{apx:forced-aligner}

\subsection{Why Do We Need a Forced Aligner?}

Modern ASR systems often output punctuated transcripts, either directly or through auxiliary punctuation restoration modules. This punctuation is usually optimized for readability and semantic plausibility, not for the acoustic pause structure of speech. As a result, ASR punctuation may correlate only weakly with real pauses, hesitations, or phrase boundaries in the waveform.

This mismatch is important when ASR-generated annotations are used to train downstream TTS systems. If punctuation does not reliably correspond to acoustic pauses, a TTS model may learn weak or inconsistent pause control: punctuation may fail to trigger a pause, while pauses may appear where no punctuation exists. These errors degrade downstream TTS prosody and controllability.

This motivates a dedicated forced aligner that grounds textual units in the speech signal and recovers word boundaries and pause structure from acoustic evidence rather than ASR punctuation conventions.

For expressive in-the-wild audio, raw recordings are rarely useful as training supervision without reliable transcripts, temporal boundaries, and fine-grained attribute labels. Otherwise, annotation errors are inherited by downstream generation models~\cite{zhang2024gtsinger,li2024robust,guo2025stars}. The problem becomes more pronounced in structured controllable audio generation, where the model must separate linguistic content, speaker identity, pronunciation, style, and expressive factors from imperfect supervision in large-scale real-world pipelines~\cite{zhang2024stylesinger,zhang2024tcsinger,zhang-etal-2025-tcsinger,guo2025techsinger}.

\subsection{Overview}

Forced alignment aligns a transcript with a speech waveform and predicts temporal boundaries such as word-level start and end times. In practice, pauses, variable speaking rates, weak articulations, and annotation noise, including zero-duration or near-zero-duration labels, can degrade alignment quality. These issues are more difficult for aligners that rely on a single global blank representation or purely local frame classification without explicit sequence-structure control or learned transition constraints.

\begin{itemize}
    \item Traditional forced aligners such as Montreal Forced Aligner (MFA)\cite{mcauliffe2017montreal} rely on pronunciation lexicons and Kaldi-style triphone acoustic modeling with speaker adaptation. They remain strong and widely used baselines, especially in lexicon-rich settings. Their modeling assumptions, however, make it less direct to add task-specific neural representations, structured blank modeling, and learned transition preferences.
    \item A related line of work treats alignment or segmentation as frame-level boundary classification or segmentation \cite{zhu2022phone,strgar2023phoneme}, which is especially relevant for phone-level alignment and boundary-sensitive tasks. Boundary detection and frame-wise classification, however, do not by themselves define a globally consistent word-to-speech alignment path. Transcript-conditioned word-level alignment often needs additional mechanisms to enforce monotonic occupancy, represent heterogeneous gap states, and stabilize ambiguous cases.
    \item CTC-based systems \cite{rastorgueva2023nemo} and ASR-alignment pipelines such as WhisperX \cite{bain2023whisperx} obtain timestamps from implicit CTC paths or auxiliary alignment stages. This works well as an engineering pipeline, but pause regions, blank handling, and transition preferences are spread across separate components rather than learned in one transcript-conditioned objective.
    \item Recent methods such as Canary \cite{hu2025word} and Qwen3-Omni \cite{xu2025qwen3} use large-scale neural models to predict timestamps directly. These models are typically large and autoregressive, which can be expensive for large-scale offline processing and online lyric/subtitle alignment services. Concurrent work, Qwen3 Forced Aligner \cite{mu2026llm}, proposes a non-autoregressive parallel slot-filling approach that also leverages multilingual semantic knowledge from pretrained large language models. Its design concatenates audio features, text, and time slots into one sequence, so activation and memory cost scale with the joint sequence length.
\end{itemize}

We focus on transcript-conditioned word-level forced alignment, especially when downstream speech generation or annotation refinement requires accurate pause-aware boundaries. Swan Forced Aligner combines (i) an explicit interleaved word/blank state topology, (ii) structured decoding with calibrated unary and transition scores, and (iii) an optional posterior-based decoding mode for locally ambiguous evidence in noisy long-form speech segments.

Compared with direct timestamp prediction, our model maintains an explicit alignment lattice with monotonic structural constraints, making the decoding process more controllable, interpretable, and diagnosable. The model is also computationally efficient, with compact parameterization, modest activation memory, and low-latency Viterbi decoding. Compared with conventional frame-classification aligners, it models both state emissions and state transitions in one alignment framework, and supports topology-constrained posterior decoding under uncertain or weak acoustic evidence.

In long-form structured audio modeling, small local timing errors can accumulate into content drift, unstable conditioning, or mismatches across the generated audio. Accurate time structure is therefore a practical requirement rather than a cosmetic annotation detail~\cite{zhang2025conan,zhang2025versband,guo2025mrsaudio}. These difficulties also motivate broader evaluation protocols, since perceived quality depends on frame-level fidelity, consistency, preference, and synchronization over longer temporal contexts~\cite{zhu2025asaudio,pan2025mesa}.

\begin{figure}[!htbp]
  \centering
    \includegraphics[width=0.8\textwidth]{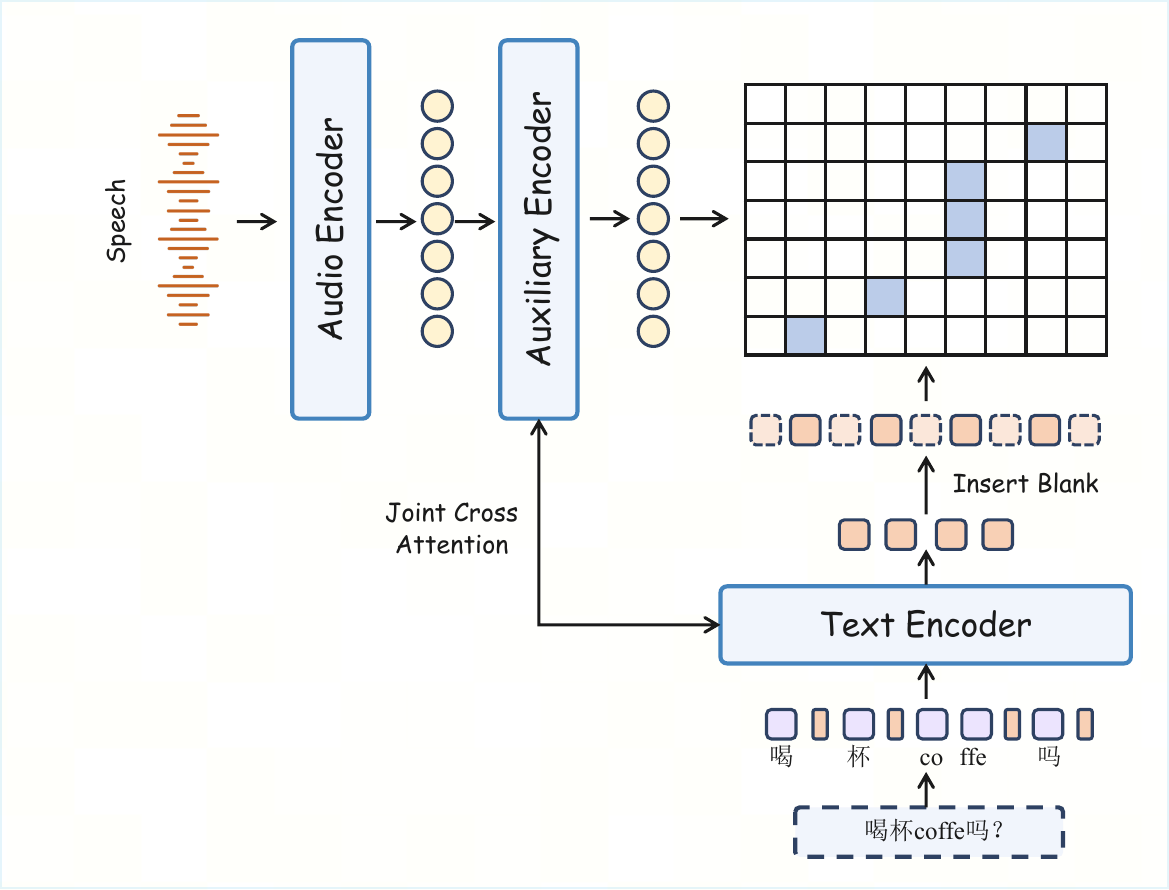}
  \caption{Overview of Swan Forced Aligner.}
  \label{fig:aligner}
\end{figure}

\section{Method}

\subsection{Problem Setup}

Let $x$ denote an input speech waveform and let $y=(y_1,\dots,y_N)$ denote its transcript. Our goal is to estimate the temporal boundary of each word in the transcript, i.e., a sequence of word-level intervals
\[
\{(s_i, e_i)\}_{i=1}^M,
\]
where $M$ is the number of aligned lexical words, and $s_i$ and $e_i$ denote, respectively, the start and end times of the $i$-th word on the input waveform time axis.

We focus on transcript-conditioned word-level forced alignment. The transcript is assumed to be given, so the main challenge is not lexical recognition but robust boundary localization under pauses, speaking-rate variation, weak articulations, and annotation noise. Realistic supervision may contain uncertain boundaries, heterogeneous blank regions between adjacent words, and zero-duration labels.

For training, each utterance may optionally be associated with word-level annotations
\[
\{(\hat{s}_i, \hat{e}_i)\}_{i=1}^M,
\]
and, when available, a confidence score $\hat{c}_i \in [0,1]$ for each word annotation. These annotations are used to construct frame-level occupancy targets and duration supervision.

\subsection{Lexical Word Representation}

The transcript is represented as a sequence of lexical word units,
\[
g = (g_1, \dots, g_M),
\]
where each $g_i$ denotes one word to be aligned. Each word is then tokenized by a predefined text tokenizer into one or more subword tokens. The tokenizer used by SwanVoice does not merge multiple lexical words into a single token, but it may split a word into multiple tokens, especially for English. For example, a Chinese character is typically mapped to one token, while an English word may be decomposed into several subword pieces.

This tokenizer behavior is convenient for text modeling, but it creates a granularity mismatch for word-level forced alignment: the alignment target is a lexical word, whereas the text encoder operates on tokenizer-level units. We bridge this gap by inserting a dedicated anchor symbol \texttt{<|wbd|>} after each lexical word. Denote the tokenizer output of $g_i$ as
\[
B(g_i) = (t_{i,1}, \dots, t_{i,n_i}),
\]
The final token sequence is
\[
\tilde{y} =
(t_{1,1}, \dots, t_{1,n_1}, \texttt{<|wbd|>},
 \dots,
 t_{M,1}, \dots, t_{M,n_M}, \texttt{<|wbd|>}).
\]

The special token \texttt{<|wbd|>} acts as a word-level alignment anchor. It aggregates the contextual information of the preceding subword span into one hidden state representing the lexical word. Swan Forced Aligner therefore aligns word-anchor states extracted from the contextualized hidden states at \texttt{<|wbd|>} positions, rather than aligning every subword token independently.

\subsection{Backbone Encoders}

A pretrained acoustic encoder maps the input waveform $x$ to frame-level features:
\[
A^{(0)} = \mathrm{Enc}_{\mathrm{aud}}^{\mathrm{pre}}(x) \in \mathbb{R}^{T \times d_a},
\]
where $T$ is the number of acoustic frames after subsampling and $d_a$ is the hidden dimension of the pretrained encoder. A lightweight Transformer encoder refines these projected features:
\[
A = \mathrm{Enc}_{\mathrm{aud}}(\mathrm{Proj}_{\mathrm{aud}}(A^{(0)})) \in \mathbb{R}^{T \times d}.
\]
This yields the frame-level acoustic sequence $A=(a_1,\dots,a_T)$ used by the structured aligner.

On the text side, the text encoder maps the tokenized sequence $\tilde{y}$ to contextualized representations:
\[
H = \mathrm{Enc}_{\mathrm{text}}(\mathrm{Embed}(\tilde{y})) \in \mathbb{R}^{L \times d}.
\]

The acoustic and text streams are not fully independent. The backbone allows text-conditioned acoustic encoding and audio-conditioned text encoding, so the resulting representations already carry cross-modal alignment cues before structured decoding.

Gathering the hidden states at the \texttt{<|wbd|>} positions gives the word-level text-anchor sequence:
\[
W = (w_1,\dots,w_M), \ w_i \in \mathbb{R}^d.
\]
Each anchor $w_i$ summarizes the full token span associated with lexical word $g_i$, including the case where that word is decomposed into multiple subword tokens. These word-anchor representations are used as the text-side word states in the structured aligner.

\subsection{Structured Alignment Topology}

Swan Forced Aligner performs alignment over an explicit interleaved word--blank topology rather than predicting timestamps directly from a flat sequence representation. For a transcript with $M$ lexical words, the latent state space is
\[
\mathcal{S} = (b_0, w_1, b_1, w_2, \dots, w_M, b_M),
\]
where $w_i$ denotes the $i$-th word state and $b_i$ denotes the blank or gap state before, between, or after words. This topology represents both word occupancy and the pauses, silences, and transitional blank regions that appear in real speech.

Each word state $w_i$ is represented by the corresponding word-anchor embedding from the text encoder. For blank states, Swan Forced Aligner avoids a single global blank representation and models heterogeneous blank regions explicitly. Separate learnable parameters are used for the utterance-initial blank state $b_0$ and the utterance-final blank state $b_M$. For an internal blank between adjacent words, the prototype is conditioned on both neighboring word states:
\[
b_i = b_{\mathrm{base}} + \Delta(w_i, w_{i+1}), \ 1 \le i \le M-1,
\]
where $b_{\mathrm{base}}$ is a global blank prototype and $\Delta(\cdot,\cdot)$ is a small neural module that predicts a gap-specific residual from the adjacent word-state pair. This allows the model to distinguish short coarticulatory gaps, long pauses, and phrase-level boundaries.

A valid alignment path is represented by a latent-state sequence:
\[
z = (z_1,\dots,z_T), \ z_t \in \mathcal{S},
\]
The path follows the interleaved topology monotonically and uses three transition types:
\[
\texttt{stay}, \ \texttt{adv1}, \ \texttt{adv2}.
\]
Here, \texttt{stay} keeps the current state, \texttt{adv1} advances by one state along the topology, and \texttt{adv2} skips across an intermediate blank when transitioning into a word state. These transitions enforce monotonic decoding that remains consistent with the transcript.

\subsection{State Scoring and Stability-Oriented Calibration}

Given acoustic frame features $A=(a_1,\dots,a_T)$ and state representations in $\mathcal{S}$, the model computes a frame-level unary score for each valid frame--state pair. Let $h_s \in \mathbb{R}^d$ denote the representation of state $s$. The raw unary score at frame $t$ for state $s$ is defined as
\[
u_{t,s} = \phi(a_t, h_s),
\]
where $\phi(\cdot,\cdot)$ is either cosine similarity or dot-product similarity.

In addition to frame-level state evidence, the model scores transition preferences between neighboring states. Transition scores are parameterized by lightweight neural heads conditioned on the destination state, with an additional pairwise module for skip transitions into word states. The incoming transition score for destination state $s$ and transition type $r$ is
\[
\tau(s, r), \ r \in \{\texttt{stay}, \texttt{adv1}, \texttt{adv2}\}.
\]
This allows the model to score which state is locally plausible at a frame and how likely different monotonic advances are under the current alignment context.

One practical goal of Swan Forced Aligner is stable structured decoding across machines and execution environments. In our experiments, even with deterministic controls enabled, small numerical differences can alter the decoded Viterbi path when unary and transition scores are poorly calibrated. Score canonicalization and decoupled scaling make decoding robust to numerical noise.

For unary scores, we perform per-frame canonicalization over valid states:
\[
\tilde{u}_{t,s} = \mathrm{Canon}_u(u_{t,s}),
\]
where the normalization is applied only over valid states at frame $t$. This removes sample-dependent score offset and scale variation and makes the relative ordering among candidate states more stable.

For transition scores, the same canonicalization is applied to all valid transition entries in the sample:
\[
\tilde{\tau}(s,r) = \mathrm{Canon}_{\tau}(\tau(s,r)).
\]
This reduces variation in transition magnitude across utterances and prevents the decoder from becoming overly sensitive to implementation-dependent score scales.

Finally, we use separate learnable gains for unary and transition terms:
\[
u^{*}_{t,s} = \gamma_u \, \tilde{u}_{t,s}, \quad
\tau^{*}(s,r) = \gamma_{\tau} \, \tilde{\tau}(s,r),
\]
where $\gamma_u$ and $\gamma_{\tau}$ are independent learnable parameters. This dual-gamma design is more flexible than a single global temperature because occupancy and transition terms require separate calibration.

The final score of a valid state sequence $z=(z_1,\dots,z_T)$ is defined as
\[
\mathrm{Score}(z)
=
\sum_{t=1}^{T} u^{*}_{t,z_t}
+
\sum_{t=2}^{T} \tau^{*}(z_t, r_t),
\]
where $r_t$ denotes the transition type used to enter state $z_t$ from $z_{t-1}$. The calibrated unary and transition terms define the final transcript-conditioned alignment score.

\subsection{Training Objectives}

During training, word-level time annotations are converted into frame-level state supervision over the interleaved topology. Frames assigned to lexical words are supervised by their corresponding word states, while the remaining valid frames are assigned to blank states according to their positions relative to neighboring words. This yields targets on the inference lattice.

The primary frame-level supervision is a cross-entropy alignment loss over valid acoustic frames:
\[
\mathcal{L}_{\mathrm{ce}}
=
\frac{1}{|\Omega|}
\sum_{t \in \Omega}
\alpha_t \,
\mathrm{CE}(p_t, \hat{z}_t),
\]
where $\Omega$ is the set of valid acoustic frames, $\hat{z}_t$ is the target state at frame $t$, $p_t$ is the predicted state distribution, and $\alpha_t$ is an optional frame weight derived from annotation confidence.

To encourage globally consistent alignment paths, Swan Forced Aligner also optimizes a CRF objective over the same structured lattice. Let $\mathcal{Z}$ denote the set of all valid monotonic state paths and let $\mathrm{Score}(z)$ denote the path score defined in the previous subsection. For a target path $\hat{z}$ derived from word-level time annotations, we use the CRF loss
\[
\mathcal{L}_{\mathrm{crf}}
=
-
\log
\frac{\exp(\mathrm{Score}(\hat{z}))}
{\sum_{z \in \mathcal{Z}} \exp(\mathrm{Score}(z))}.
\]
This objective complements frame-level cross-entropy by encouraging the gold alignment path to receive a high global score relative to all other valid monotonic paths.

To regularize state occupancy, Swan Forced Aligner uses duration supervision for both word states and blank states. Let $\hat{d}^{(w)}_i$ and $\hat{d}^{(b)}_i$ denote the target durations of word and blank states, and let $d^{(w)}_i$ and $d^{(b)}_i$ denote the predicted occupancies obtained by summing state posteriors over time. The two duration terms are combined as
\[
\mathcal{L}_{\mathrm{dur}}
=
\mathcal{L}_{\mathrm{dur}}^{(w)}
+
\lambda_b \mathcal{L}_{\mathrm{dur}}^{(b)},
\]
where both terms combine absolute-error and log-duration penalties to stabilize supervision across short and long segments during training on heterogeneous speech.

Swan Forced Aligner also includes a monotonicity regularization term that penalizes decreases in the expected word index over time. This encourages the word-state posterior mass to progress monotonically along the transcript and discourages locally inconsistent alignments under ambiguous evidence. Denoting this term by $\mathcal{L}_{\mathrm{mono}}$, the final training objective is
\[
\mathcal{L}
=
\mathcal{L}_{\mathrm{ce}}
+
\mathcal{L}_{\mathrm{crf}}
+
\lambda_d \mathcal{L}_{\mathrm{dur}}
+
\lambda_m \mathcal{L}_{\mathrm{mono}}.
\]
In our implementation, all major loss terms are combined with unit weight unless otherwise specified.

\subsection{Inference Procedure}

At inference time, Swan Forced Aligner computes calibrated unary and transition scores over the interleaved alignment lattice. The same monotonic topology is used for training and decoding.

The default decoding mode is Viterbi decoding, which finds the highest-scoring valid state path
\[
z^{*} = \arg\max_{z \in \mathcal{Z}} \mathrm{Score}(z),
\]
where $\mathcal{Z}$ denotes the set of all valid monotonic paths. The word-level start and end times are then recovered from the frame ranges assigned to each word state.

Swan Forced Aligner also supports an optional posterior-based decoding mode. In this mode, forward--backward inference first computes state posteriors on the same structured lattice, and a topology-constrained path is decoded using posterior scores instead of raw path scores. This mode is more robust when local evidence is ambiguous because it incorporates path uncertainty rather than relying only on a single maximum-score explanation.

After decoding, the confidence of each aligned word is estimated by aggregating emission-state probabilities over the frames assigned to that word state. Once a word-aligned frame span is determined by the decoded path, the confidence score is computed as the average word-state probability over that span. The decoded path may depend on the inference mode, while the confidence itself is still derived from emission-side state probabilities.

Together with the explicit state path, these state probabilities provide diagnostic signals for downstream debugging and alignment-error analysis.

\section{Experiments}

\subsection{Experimental Setup}

\paragraph{Datasets}
For the forced-aligner experiments in this appendix, we use a separate 80K-hour Chinese-English alignment-training subset from internal resources. It spans audiobooks, podcasts, conversational speech, meetings, and live stream recordings.
All training sets are pre-annotated with pseudo-timestamps using the Montreal Forced Aligner~(MFA). 
For evaluation, we use two human-timestamped sets: the Chinese subset of GTSinger-Speech~\cite{zhang2024gtsinger} and Librispeech-Alignment~\cite{amodei2016deep}.

\paragraph{Implementation Details}

We use WavLM\cite{chen2022wavlm} as the pretrained audio encoder. The auxiliary encoder is a 4-layer bidirectional Transformer with hidden size 512 and 8 attention heads. The text encoder is a 16-layer bidirectional Transformer with hidden size 512 and 8 attention heads. The model has about 400M parameters. Swan Forced Aligner is trained on the 80K-hour subset using 24 A100 GPUs, with a batch size of 4 hours for 80K steps. We optimize with AdamW, using a learning rate of 1.0e-5 and $\beta=(0.9,0.999)$.

\paragraph{Evaluation Metrics}

We evaluate timestamp prediction with accumulated averaging shift (AAS), following prior work~\cite{shi2022achieving}. Lower AAS indicates more accurate timestamp prediction. AAS measures the average boundary deviation across all evaluated word slots:
\begin{align}
    AAS = \dfrac{1}{N} \sum_{i=1}^N \|s_i - \hat{s}_i\|_1 = \dfrac{1}{N} \sum_{i=1}^N (|t_\text{start}^{(i)} - \hat{t}_\text{start}^{(i)}| + |t_\text{end}^{(i)} - \hat{t}_\text{end}^{(i)}|),
\end{align}
where $N$ is the number of evaluated word slots, $s_i=(t_\text{start}^{(i)},t_\text{end}^{(i)})$ is the ground-truth boundary pair, and $\hat{s}_i=(\hat{t}_\text{start}^{(i)},\hat{t}_\text{end}^{(i)})$ is the predicted boundary pair.

\paragraph{Baselines}

We compare with five mainstream forced aligners:
 (1) Monotonic-Aligner~\cite{shi2022achieving}, a non-autoregressive Paraformer-based aligner using a continuous integrate-and-fire mechanism, which supports only Chinese~\footnote{\url{https://modelscope.cn/models/iic/speech_timestamp_prediction-v1-16k-offline}}.
 (2) NeMo Forced Aligner~\cite{rastorgueva2023nemo}, a tool for
generating token-, word-, and segment-level timestamps of speech in audio using NeMo's CTC-based ASR models. We use the official checkpoint~\footnote{\url{https://ngc.nvidia.com/models/nvidia:stt_en_fastconformer_hybrid_large_pc}} to perform English alignment. (3) WhisperX~\cite{bain2023whisperx}, a time-accurate speech recognition system with word-level timestamps utilizing voice activity detection and forced phoneme alignment. We use different checkpoints for Chinese and English speech following the official inference script~\footnote{\url{https://github.com/m-bain/whisperX/blob/main/whisperx/alignment.py}}; (4) Qwen3 Forced Aligner~\cite{mu2026llm}, a non-autoregressive aligner based on parallel slot filling and multilingual speech-language representations. We perform alignment using their official checkpoint~\footnote{\url{https://github.com/QwenLM/Qwen3-ASR}}; (5) LattifAI Aligner, a speech agent for millisecond-precision alignment. We use their official SDK~\footnote{\url{https://github.com/lattifai/lattifai-python}} and the released \texttt{Lattice-1} checkpoint~\footnote{\url{https://huggingface.co/LattifAI/Lattice-1}} across all evaluation runs.

\subsection{Experimental Results}
\begin{table*}[htp]
\caption{AAS (ms)$\downarrow$ of \textbf{Swan Forced Aligner} and other forced aligners on Chinese and English test datasets. The best results are in \textbf{bold} and the second best are \underline{underlined}. * denotes checkpoints that were not publicly released at evaluation time.}
\centering
\begin{tabular}{@{}cccc@{}}
\toprule
                    & \multicolumn{1}{c}{GTSinger-Speech-ZH} & \multicolumn{1}{c}{LibriSpeech-Clean} & \multicolumn{1}{c}{LibriSpeech-Others} \\ \midrule
Monotonic-Aligner   & 61.98                           & -                                     & -                                      \\
NeMo Forced Aligner & -                               & 87.05                                 & 91.85                                  \\
WhisperX            & 221.29                          & 87.02                                 & 96.64                                  \\
Qwen3 Forced Aligner & 47.31                          & 27.84                                 & \textbf{29.74}                         \\
LattifAI Aligner*   & \textbf{31.60}                  & \textbf{25.70}                        & 36.00                                  \\ \midrule
Swan Forced Aligner        & \underline{45.19}          & \underline{27.67}                    & \underline{29.92}                     \\ \bottomrule
\end{tabular}
\label{tab:aligner}
\end{table*}

As shown in Table~\ref{tab:aligner}, Swan Forced Aligner gives the best open-source AAS on the Chinese and LibriSpeech-Clean benchmarks. On LibriSpeech-Others, it is within 0.18 ms of Qwen3 Forced Aligner and about 10 ms behind LattifAI Aligner, the best proprietary system in this comparison.

\end{document}